\documentclass[12pt]{article}
\usepackage{amsmath,amssymb,fullpage}
\usepackage{natbib}

\begin{document}
\begin{center}
\textbf{\large{Discussion on Bayesian Cluster Analysis: Point Estimation and Credible Balls by Sara Wade and Zoubin Ghahramani}}\\ \vspace{10pt}
\textbf{William Weimin Yoo}\\
\textit{Leiden University}
\end{center}
\begin{abstract}
I begin my discussion by giving an overview of the main results. Then I proceed to touch upon issues about whether the credible ball constructed can be interpreted as a confidence ball, suggestions on reducing computational costs, and posterior consistency or contraction rates.
\end{abstract}

\vspace{10pt}\\

The authors should be congratulated for producing such an interesting and important work. In the present paper, \citet{wade2018} investigated the issues of point estimation and uncertainty quantification for Bayesian clustering analysis. Here, the data density is modelled as a countably infinite mixture and latent variables attaching to each observation are introduced to represent cluster membership. A common prior for the mixing distribution is the Dirichlet process, and they used this as the default prior in the simulations and real data analysis. They derived point estimators through decision theory by considering two different clustering losses/metrics, i.e., Binder's loss ($N$-invariant version) and variation of information (VI). They endowed the space of partitions with a lattice by including partial order and the covering relation, and this enables them to compare properties of these two metrics and define a consistent notion of closeness between partitions. This latter notion was further used to develop a method to construct credible ball over partitions using the aforementioned metrics. The optimization problem needed to find the point estimate (for VI) is computational demanding and the search space is very high-dimensional. To scale up computations, the authors proposed a greedy search algorithm.

I start my discussion by asking the question whether the credible balls constructed can be interpreted as confidence balls in the frequentist sense? Specifically, do the $95\%$ credible balls based on Binder's loss or VI with their vertical and horizontal bounds, have also approximate $95\%$ frequentist coverage probability (contains the true clustering $95\%$ of the time)? For finite dimensional parameters, we have the Bernstein-von Mises theorem to ensure this equivalence; however in the nonparametric setting as in this paper, this equivalence breaks down and it is in general not true that Bayesian credible ball is also a frequentist confidence ball. It would be very interesting if we can give some theoretical guarantees on coverage for the VI credible ball, or maybe compare the extent of its uncertainty in a simulation with a confidence ball over partitions constructed based on non-Bayesian methods (if there are any). In complex models, it is straightforward to use Markov Chain Monte Carlo (MCMC) samples to construct credible balls, as compared to frequentist methods which rely on complicated asymptotic normality analysis or bootstrap, and hence such comparisons and coverage guarantees will provide good incentives for statisticians (particularly non-Bayesians) to use the methods proposed in this paper to do clustering in their own work.

A recurring theme that came up when designing algorithms in the paper is the ability to scale to massive datasets and to speed up computations. Instead of using infinite mixtures which entails searching over the entire partition space, one can use overfitted mixtures as investigated in \citet{rousseau2011}, where one intentionally overfit the model by choosing a larger but finite number of components than necessary and use some sparsity-inducing priors to zero out the unnecessary components. Alternatively, by observing in Table 2 that the number of clusters for the VI credible ball stays constant for the different sample sizes considered, its robust property suggests that we could first try to estimate the correct number of clusters, through MAP (Maximum a posteriori) or the recently proposed Bayes Lepski's method (\citealp{yoo2018}), and only explore the part of the partition space corresponding to this estimated number of clusters.

I totally agree with the authors that we need results on posterior consistency and contraction rates, in order to fully resolve the ambiguity caused by the positive results of the present paper and the negative results of \citet{miller2014}. Question of interests include characterizing the rate at which the number of clusters estimated under the VI posterior approaches the true number, and whether this rate is optimal. In addition, it would also be interesting to study miss-classification errors and how they grow with sample size or depend on the chosen loss function. A deeper understanding of these issues will help statisticians choose the right priors and design algorithms to control these errors.

The present paper proposes a very promising method to obtain point estimate and uncertainty quantification for Bayesian cluster analysis, which is a great improvement in terms of interpretability over posterior similarity matrices commonly considered in the literature. I envision that the lattice-based framework introduced here can be extended to other settings as well, e.g., multiple membership clusters, and I am certain this work will further spur research in these areas.

\bibliographystyle{apa}
\bibliography{cluster}

\end{document}